# A chemical study of nine star-forming regions with evidence of infall motion

Yang Yang,[1,2,3]★ Yao Wang,[1]★ Zhibo Jiang[1,4]★ and Zhiwei Chen[1]

[1]*Purple Mountain Observatory, Chinese Academy of Sciences, Nanjing 210023, China*
[2]*University of Science and Technology of China, Chinese Academy of Sciences, Hefei 230026, China*
[3]*Center for Astrophysics, Guangzhou University, Guangzhou 510006, China*
[4]*Center Astronomy and Space Sciences, China Three Gorges University, Yichang 443002, China*

**ABSTRACT**
The study of the physical and chemical properties of gas infall motion in the molecular clumps helps us understand the initial stages of star formation. We used the FTS wide-sideband mode of the IRAM 30-m telescope to observe nine infall sources with significant double-peaked blue line profile. The observation frequency ranges are 83.7–91.5 GHz and 99.4–107.2 GHz. We have obtained numbers of molecular line data. Using XCLASS, a total of 7–27 different molecules and isotopic transition lines have been identified in these nine sources, including carbon chain molecules, such as CCH, c-$C_3H_2$ and $HC_3N$. According to the radiation transfer model, we estimated the rotation temperatures and column densities of these sources. Chemical simulations adopting a physical model of HMSFRs are used to fit the observed molecular abundances. The comparison shows that most sources are in the early HMPO stage, with the inner temperature around several 10 K.

## 1 INTRODUCTION

Protostars are formed through the gas inside-out gravitational collapse in the dense core (e.g. Shu, Adams & Lizano 1987). Gas infall motion mainly occurs at the beginning of the star formation process (e.g. Bachiller 1996), and will continue for a period of time to accumulate the material. Therefore, studying the gas infall motion will help us better understand the initial stages of star formation. However, it is difficult to observe infall motion directly. According to the model (e.g. Leung & Brown 1977), the gas infall motion causes a redshifted self-absorption profile for the optically thick line, resulting in an asymmetric double-peaked profile with a blue peak stronger than the red peak, commonly named as the blue profile (e.g. Mardones et al. 1997). Therefore, we generally use the blue profile of optically thick lines produced by infalling gas as indirect evidence to search and identify the infall candidates. However, in some cases, multiple velocity components can also cause similar double-peaked profiles. This requires optically thin line data to trace the central radial velocity of the clumps. If the optically thin line shows a single-peaked profile located between the double peaks of the optically thick line, the blue profile is more likely due to the gas infall motion. Therefore, the combination of optically thick and thin lines is essential in the identification of infall candidates. Among them, the commonly used optically thick lines are $HCO^+$ (1-0, 3-2, 4-3), $^{12}CO$ (2-1, 3-2, 4-3), HNC (1-0, 4-3), HCN (1-0, 3-2), CS (2-1), and $H_2CO$ ($2_{12}$-$1_{11}$) etc., while $N_2H^+$ (1-0), $C^{18}O$ (1-0, 2-1, 3-2), $H^{13}CO^+$ (1-0), and $C^{17}O$ (3-2) are usually used as optically thin lines (e.g. Yu et al. 2022, RAA).

Some observations suggest that, except of gravitational collapse, there are other star-forming activities in the infall source. Numerous observations indicated that molecular outflows also exist in infall sources (e.g. Zhu, Zhao & Wright 2011; Klaassen, Testi & Beuther 2012; Kim, Kim & Kim 2015; Zhang et al. 2017; Li et al. 2019). In some cases, the gas infall motion is coupled with the outflow, increasing the classification ambiguity for the spectral line profile of single-point observation, and usually requires mapping observations to provide more spatial details (e.g. He et al. 2016; Juárez et al. 2019). Other studies have detected abundant masers in the infall sources, such as methanol masers and water masers (e.g. Schneider et al. 2010; Li et al. 2019). Moreover, abundant molecules were detected in some infall sources. Using the 20-m radio telescope of the Onsala Space Observatory (Sweden), Pirogov et al. (2016) observed five massive star-forming regions, including two infall candidates, 121.28 + 0.65 and 99.982 + 4.17. They observed a number of molecular spectral lines in the frequency range of 85–89 GHz, including some carbon-chain species such as c-$C_3H_2$, $CH_3CCH$, and CCH. Analysing observation results of these molecules and calculating their abundance ratios can help us to roughly estimate the evolutionary time-scale of infall sources. Mookerjea et al. (2012) used the Herschel's HIFI and the IRAM 30-m radio telescope to observe $C_3$, c-$C_3H_2$, and CCH lines in an infall source DR21(OH). They compared the molecular abundances calculated from $C_3$, c-$C_3H_2$ and CCH (87 and 524 GHz) lines data with those simulated by chemical models adopting the Ohio State University gas–grain code with a warm-up stage (Hasegawa, Herbst & Leung 1992; Garrod & Herbst 2006). The results indicate that the evolutionary time-scale of DR21(OH) is approximately 0.7–3 Myr with the temperature around 30 K. The analysis of the comparison between chemical simulations and observations is an important method to investigate the evolutionary stage of sources, such as infall candidates.

★ E-mail: yangy@gzhu.edu.cn (YY); wangyao@pmo.ac.cn (YW); zbjiang@pmo.ac.cn (ZJ)

**Table 1.** Source list.

| Source Name | RA (J2000) | Dec. (J2000) | Vlsr (km s$^{-1}$) | Distance (kpc) | $V_{in}$ (km s$^{-1}$) | $\dot{M}_{in}$ ($\times 10^{-4} M_\odot$ yr$^{-1}$) | Association | RMS (K) | Extent of map | Number of molecules |
|---|---|---|---|---|---|---|---|---|---|---|
| G029.60-0.63 | 18:47:36.1 | −03:15:09 | 76.52 | 4.31 | 1.6(0.9) | 55 | | 0.18 | 3 × 3 arcmin$^2$ | 9 |
| G053.13+0.09 | 19:29:11.8 | +17:56:19 | 21.71 | 1.67 | 0.7(0.5) | 2.2 | IRAS 19270+1750 | 0.15 | 4 × 6 arcmin$^2$ | 16 |
| G081.72+0.57 | 20:39:01.0 | +42:22:41 | − 3.22 | 1.50(0.08)$^a$ | 0.4(0.6) | 22 | DR21(OH), W75 | 0.13 | 2.5 × 2.5 arcmin$^2$ | 27 |
| G082.21-1.53 | 20:49:30.3 | +41:27:35 | 3.07 | 1.13 | 0.1(0.3) | 0.57 | | 0.14 | 3 × 3 arcmin$^2$ | 9 |
| G107.50+4.47 | 22:28:32.1 | +62:58:40 | − 1.27 | 0.25$^b$ | 0.3(0.2) | 0.49 | IRAS 22267+6244 | 0.14 | 3 × 3 arcmin$^2$ | 17 |
| G109.00+2.73 | 22:47:17.6 | +62:11:44 | − 10.29 | 0.79 | 0.3(0.8) | 1.0 | | 0.15 | 3.5 × 3.5 arcmin$^2$ | 11 |
| G121.31+0.64 | 00:36:54.5 | +63:27:57 | − 17.54 | 0.89(0.05)$^c$ | 0.3(0.9) | 1.2 | IRAS 00338+6312 | 0.18 | 3 × 3 arcmin$^2$ | 16 |
| G193.01+0.14 | 06:14:25.1 | +17:43:12 | 7.59 | 1.91 | 1.5(0.8) | 11 | | 0.12 | 3 × 3 arcmin$^2$ | 13 |
| G217.30-0.05 | 06:59:14.3 | −03:54:35 | 26.83 | 2.34 | 0.2(0.1) | 2.6 | IRAS 06567-0350 | 0.13 | 2 × 2 arcmin$^2$ | 7 |

The distance of the sources are obtained from: $^a$Rygl et al. (2012); $^b$Bailer-Jones et al. (2018); $^c$Rygl et al. (2008), where $^a$ and $^c$ are the parallax distances. The $V_{in}$ and $\dot{M}_{in}$ values are adopted from Yang et al. (2021).

In previous studies, we used the Fourier transform spectrometer (FTS) wide-sideband mode of the IRAM 30-m telescope to map a sample of infall sources, and obtained multiple molecular line data. These data can be used to analyse the physical and chemical properties of infall sources. The targets we observed were selected from the $^{12}$CO (1-0) and its isotopes $^{13}$CO (1-0) and C$^{18}$O (1-0) data of the Milky Way Imaging Scroll Painting (MWISP) project (Jiang et al., in preparation). To confirm these candidates, we used the HCO$^+$ (1-0) line, which is more generally used to trace infall, to make single-point and mapping observations of the infall candidates. The Purple Mountain Observatory 13.7-metre telescope was used for single-point observations of the candidates, and the IRAM 30-metre telescope was used for mapping observations of the sources confirmed by single-point observations (Yang et al. 2020, 2021). We have identified 19 sources with infall characteristics through the HCO$^+$ mapping observations, and nine of them show typical double-peaked blue profiles with sufficient signal-to-noise ratio.

In this paper, we firstly present the mapping observation results of these nine star-forming regions with typical infall profiles, and analyse the multiple spectral line data of them. Multiple spectral lines analysis of other infall candidates with single-peaked and peak-shoulder blue profiles will be given in the future. In Section 2, we briefly introduce the mapping observations. The lines identification and sources' morphology are given in Section 3. In Section 4.1, we use eXtended CASA Line Analysis Software Suite (XCLASS) to fit the lines and estimate their rotation temperatures, column densities, and the molecular abundances. In Section 4.2, a comparison of molecular abundances between observations and chemical simulations is presented. Finally, Section 5 summarizes the results and analysis of this work.

## 2 SOURCES AND OBSERVATION

We have used the IRAM 30-m telescope to observe some sources with infall evidence. In this paper, we will analyse the multiple spectral lines data of nine of them. All these sources show typical double-peaked blue profiles, and were identified as infall sources with higher confidence in previous study (Yang et al. 2021). These nine infall sources are listed in Table 1, where the distance values are given in column 5. We prefer to use the distances given in the literature, especially the more reliable values estimated by the triangular parallax method. If the source's distance is not given in the literature, we use the Reid et al. (2014) model to estimate its kinematic distance. The gas infall velocities and mass infall rates of these infall sources are listed in the columns 6 and 7 of Table 1, respectively, and the data are from the table 6 of Yang et al. (2021).

The mass infall rate of these sources suggests that intermediate or massive stars may be forming in these regions.

In a previous study, we found that these infall sources also have evidence for outflow and other star-forming activities. Except for G082.21-1.53 and G109.00+2.73, bipolar or unipolar outflows have been found in these sources. Except for the two sources mentioned above and G193.01+0.14, one or more kinds of masers including water masers have been observed in the other six sources [see the section 5.2 of Yang et al. (2021)]. Water masers are usually generated in shock waves produced by outflows and stellar winds (e.g. Torrelles et al. 2005), which may indicate that these infall sources are at a later stage of collapse.

The observations were carried out in 2018 February 20–March 27 (project code: 134-18) and 2019 June 11–19 (project code: 032-19). The typical system temperature ($T_{sys}$) is between 120 and 130 K. The receiver we used is Eight Mixer Receivers and the backends are VESPA and FTS. In Yang et al. (2021), we analysed the observation results of the two spectral lines HCO$^+$ (1-0) and H$^{13}$CO$^+$ (1-0), which will not be repeated. In this paper, we mainly analyse the observation results of the multiple spectral lines obtained by the FTS. The observations cover the frequency ranges 83.7–91.5 GHz and 99.4–107.2 GHz with the frequency resolution of 195 kHz. We used the OTF PSW observing mode to cover the targets. The angular resolution of IRAM 30-m at 3-mm band is about 28 arcsec, and the main beam efficiency is approximately from 81 per cent to >78 per cent at these frequency ranges.[1] The area size of these targets depends on the MWISP $^{13}$CO (1-0) and C$^{18}$O (1-0) mapping results. Including the time spent on source integration time, pointing, focusing, and calibrating, it takes about 14 h to observe these nine sources.

The program CLASS of the GILDAS package[2] is used to reduce the raw data. The median RMS values of these sources are from 0.12 to 0.18 K at the velocity resolution of 0.67 km s$^{-1}$. We subtracted linear baselines from the spectral regions, and used the radiation transfer model to identify and fit the observed spectral lines.

## 3 RESULTS OF THE OBSERVATIONS

### 3.1 Line identification

We used the XCLASS[3] (Möller, Endres & Schilke 2017) to identify the observed lines. The spectroscopic data are taken from the Cologne

---

[1] https://publicwiki.iram.es/Iram30mEfficiencies
[2] http://www.iram.fr/IRAMFR/GILDAS/
[3] https://xclass.astro.uni-koeln.de/Home

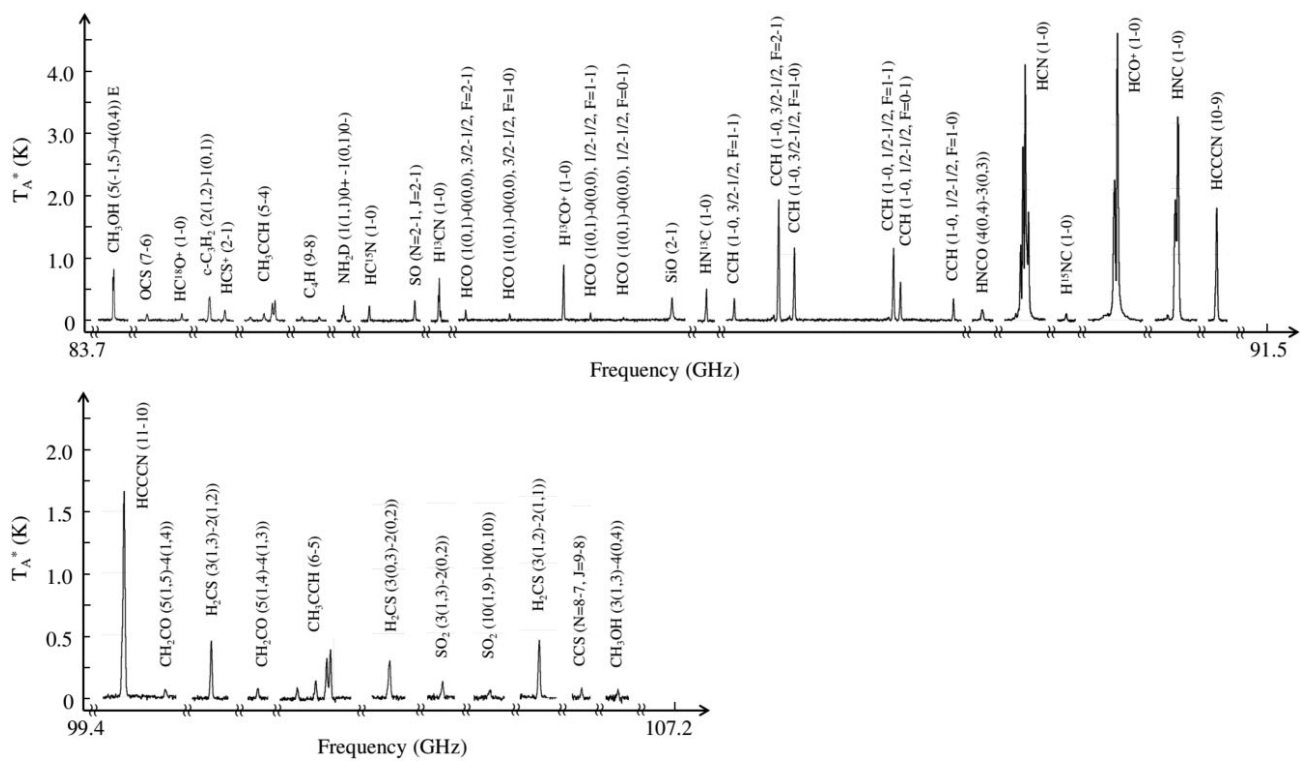

**Figure 1.** Spectrum towards the peak of the molecular emission in G081.72+0.57, observed using FTS with two 8-GHz bandwidth (lower side band and upper side band) and a 195 KHz frequency resolution. The top panel shows the spectrum of the low band and the frequency range of 83.7–91.5 GHz. The lower panel shows the spectrum of the up band and the frequency range of 99.4–107.2 GHz.

Database of Molecular Spectroscopy (Müller et al. 2001; Endres et al. 2016) and splatalogue database for astronomical spectroscopy (SPLATALOGUE).[4] We identified all emission lines with a signal-to-noise ratio greater than 3 in the average spectrum. For each line, we search for transitions with a maximum upper energy of 1000 K, and assume an excitation temperature of 20 K, a spectral line width of about 2 km s$^{-1}$, and a molecular column density of about $1 \times 10^{13}$ cm$^{-2}$. If a species produces a spectrum that roughly matches the observed lines, we identified that the spectral line is produced by that species. Fig. 1 shows the spectrum of G081.72+0.57. Shown in the figure is a part of the spectrum analyser band, divided into segments with molecular lines, and the molecular transition energy level corresponding to each emission line is also marked. We list the recognized molecular names, transition energy levels, and rest frequency in Table 2. The spectra of other targets are shown in Appendix A.

We have identified 7–27 kinds of different molecules and their isotopes emission lines, as well as a maser emission line (84.5 GHz Class I methanol maser) in these nine infall sources. These include some carbon-chain species, such as c-$C_3H_2$, CCH, $HC_3N$, and $CH_3CCH$. We detected $H^{13}CO^+$, $HCO^+$, HCN, HNC, c-$C_3H_2$, and CCH lines in all sources. Among these sources, G217.30-0.05 has the fewest molecular emission lines. More molecular lines were observed in the sources associated with massive star-forming regions, such as G081.72+0.57 and G121.31+0.64. G081.72+0.57 is associated with DR21(OH), and has the most kinds of molecular lines. Some molecules have multiple emission lines in the observed frequency ranges, such as CCH, $CH_3CCH$, $C_4H$, $NH_2D$, HCO, and $H_2CS$, allowing us to further limit the physical parameters such as excitation temperature and column density of these molecules by fit the multiple transition lines of the same molecule. We will analyse this in Section 4.1.

### 3.2 Mapping results

Maps of molecular lines observed in the nine infall sources are shown in Appendix B. Here, we show the maps of nine kinds of molecules observed in G029.60-0.63 as an example. For molecules with multiple emission lines, we only plot the integrated intensity map of the line with the highest signal-to-noise ratio. As shown in Fig. 2, the molecular name, transition energy level, and the corresponding integral velocity range are marked in the upper-left corner of the figure. The maps also show the positions of IRAS point sources and masers, as well as the positions of young stellar objects (YSOs). The maser's positions are from Yang et al. (2017, 2019), Valdettaro et al. (2001), Anglada et al. (1996), Qiao et al. (2016, 2018, 2020), etc. The YSOs are from the AllWISE source table (Cutri et al. 2013) and are classified by the YSOs criteria suggested by Koenig et al. (2012). The red pentagram symbol in the figure indicates the position of the most significant infall line profile of HCO$^+$ (1-0) identified by Yang et al. (2021).

In most cases, integrated intensity maps of different molecules will display spatially correlated compact regions, which often are associated with YSOs. Evidence of gas infall or outflow motion can also be observed in this area. In some case, the spatial distributions of some molecules are not consistent, which may be related to the difference of traced regions or the different optical depths of these lines. For G029.60-0.63, its HCN, HCO$^+$, and HNC lines are distributed in a more extended spatial region than other molecular lines. HNCO and HC$_3$N show two clumps along the northeast-

[4] https://splatalogue.online//

**Table 2.** List of observed molecular lines.

| Molecule | Transition | Frequency (MHz) | Molecule | Transition | Frequency (MHz) |
|---|---|---|---|---|---|
| $CH_3OH$ | 5(−1,5)-4(0,4) E | 84521.2 | $H^{13}CO^+$ | 1-0 | 86754.3 |
| | 3(1,3)-4(0,4) | 107013.8 | SiO | 2-1 | 86847.0 |
| OCS | 7-6 | 85139.1 | $HN^{13}C$ | 1-0 F = 0-1 | 87090.7 |
| $HC^{18}O^+$ | 1-0 | 85162.2 | | 1-0 F = 2-1 | 87090.9 |
| $HC_5N$ | 32-31 | 85201.3 | | 1-0 F = 1-1 | 87090.9 |
| | 33-32 | 87863.6 | CCH | 1-0 3/2-1/2 F = 1-1 | 87284.2 |
| $c-C_3H_2$ | 2(1,2)-1(0,1) | 85338.9 | | 1-0 3/2-1/2 F = 2-1 | 87316.9 |
| | 4(3,2)-4(2,3) | 85656.4 | | 1-0 3/2-1/2 F = 1-0 | 87328.6 |
| $HCS^+$ | 2-1 | 85347.9 | | 1-0 1/2-1/2 F = 1-1 | 87402.0 |
| $CH_3CCH$ | 5(3)-4(3) | 85442.6 | | 1-0 1/2-1/2 F = 0-1 | 87407.2 |
| | 5(2)-4(2) | 85450.8 | | 1-0 1/2-1/2 F = 1-0 | 87446.5 |
| | 5(1)-4(1) | 85455.7 | HNCO | 4(0,4)-3(0,3) | 87925.2 |
| | 5(0)-4(0) | 85457.3 | | 4(1,3)-3(1,2) | 88239.0 |
| | 6(4)-5(4) | 102516.6 | HCN | 1-0 F = 1-1 | 88630.4 |
| | 6(3)-5(3) | 102530.3 | | 1-0 F = 2-1 | 88631.8 |
| | 6(2)-5(2) | 102540.1 | | 1-0 F = 0-1 | 88633.9 |
| | 6(1)-5(1) | 102546.0 | $H^{15}NC$ | 1-0 | 88865.7 |
| | 6(0)-5(0) | 102548.0 | $HCO^+$ | 1-0 | 89188.5 |
| $C_4H$ | 9-8 J = 19/2-17/2 | 85634.0 | HNC | 1-0 F = 0-1 | 90663.5 |
| | 9-8 J = 17/2-15/2 | 85672.6 | | 1-0 F = 2-1 | 90663.6 |
| | 11-10 J = 23/2-21/2 | 104666.6 | | 1-0 F = 1-1 | 90663.7 |
| | 11-10 J = 21/2-19/2 | 104705.1 | CCS | N,J = 7,6-6,5 | 86181.4 |
| $NH_2D$ | 1(1,1)0+ -1(0,1)0- F = 0-1 | 85924.7 | | N,J = 7,7-6,6 | 90686.4 |
| | 1(1,1)0+ -1(0,1)0- F = 2-1 | 85925.7 | | N,J = 8,7-7,6 | 99866.5 |
| | 1(1,1)0+ -1(0,1)0- F = 2-2 | 85926.3 | | N,J = 8,9-7,8 | 106347.7 |
| | 1(1,1)0+ -1(0,1)0- F = 1-2 | 85926.9 | $HC_3N$ | 10-9 | 90979.0 |
| | 1(1,1)0+ -1(0,1)0- F = 1-0 | 85927.7 | | 11-10 | 100076.4 |
| $HC^{15}N$ | 1-0 | 86055.0 | $CH_2CO$ | 5(1,5)-4(1,4) | 100094.5 |
| SO | N,J = 2,2-1,1 | 86094.0 | | 5(1,4)-4(1,3) | 101981.4 |
| | N,J = 5,4-4,4 | 100029.6 | $H_2CS$ | 3(1,3)-2(1,2) | 101477.9 |
| $H^{13}CN$ | 1-0 F = 1-1 | 86338.7 | | 3(0,3)-2(0,2) | 103040.5 |
| | 1-0 F = 2-1 | 86340.2 | | 3(2,1)-2(2,0) | 103051.9 |
| | 1-0 F = 0-1 | 86342.3 | | 3(1,2)-2(1,1) | 104617.1 |
| HCO | 1(0,1)-0(0,0) 3/2-1/2 F = 2-1 | 86670.8 | $SO_2$ | 3(1,3)-2(0,2) | 104029.4 |
| | 1(0,1)-0(0,0) 3/2-1/2 F = 1-0 | 86708.4 | | 10(1,9)-10(0,10) | 104239.3 |
| | 1(0,1)-0(0,0) 1/2-1/2 F = 1-1 | 86777.4 | | | |
| | 1(0,1)-0(0,0) 1/2-1/2 F = 0-1 | 86805.8 | | | |

The spectroscopic data are from https://splatalogue.online//.

southwest direction, possibly associated with two YSOs in the figure, respectively. The confirmed infall position deviates from these two clumps, and the infall lines profiles are mainly distributed in the north–west of these two clumps, which coincides with the northerly emission area traced by $HCO^+$. For G053.13-0.09, except for $NH_2D$ and SiO, the emissions of other molecular lines are concentrated in the same area, which are basically consistent with the confirmed infall position. The spatial distributions of $NH_2D$ and SiO are slightly different. Among them, $NH_2D$ is mainly distributed in the north-east of the source, and SiO is distributed to the east of the source.

G081.72+0.57 is located in a massive star-forming region and is associated with DR21(OH). The star formation activity in this area is very active, with not only gas infall motions, but also outflow and rich maser emissions. We have observed the most kinds of molecular emission lines in this target. Among them, the spatial distributions of SO, OCS, SiO, $SO_2$, and $CH_3OH$ are more compact, while others are more extended. $NH_2D$ spatial distribution is slightly different from other molecules, and it is mainly distributed in the northern of the clump.

Another source with a complex structure is G107.50+4.47. The observation results show that both HNC, $HCO^+$ and $H^{13}CO^+$ integrated intensity maps of this source have a similar complex geometric structure. This source is likely to consist of multiple clumps. Following Yang et al. (2021), we mark these clumps as A, B, C, and D. Among them, the most obvious infall line profile was observed in clump C. These clumps are similar in velocity range, but they can still be roughly distinguished by velocity and coordinate position. We have marked the four clumps in the first map of Fig. B5. Their velocity ranges are [−4.7, −1.4], [−3.4, −1.4], [−2.8, −1.4], and [−1.0, 0.4] km s$^{-1}$, respectively. The integrated intensity maps of $H_2CS$ and HNCO are limited by insufficient signal-to-noise ratio, but it can still be seen that these two molecules are mainly distributed in the clump C. HCN and its isotope $H^{13}CN$ are mainly distributed in clump A, C, and D. CCH, $NH_2D$, $HC_3N$, $HN^{13}C$, and $c-C_3H_2$ are only observed in the clump D. Carbon chain molecules, such as $HC_3N$, are the best probes of the less chemically evolved cores, or 'chemically fresh' material (e.g. Pineda et al. 2020). The carbon chain molecules CCH, $HC_3N$, and $c-C_3H_2$ in G107.50+4.47 are all concentrated near the clump D. Although we did not observe infall line profiles on clump D, this clump also seems to be a fairly young one. In addition, SO, SiO, $SO_2$, and 84.5 GHz methanol maser are distributed in clump A and C. But there are still some differences in the spatial distribution of these molecules. SiO and $SO_2$ are mainly distributed in clump A, while SO is mainly distributed in clump C.

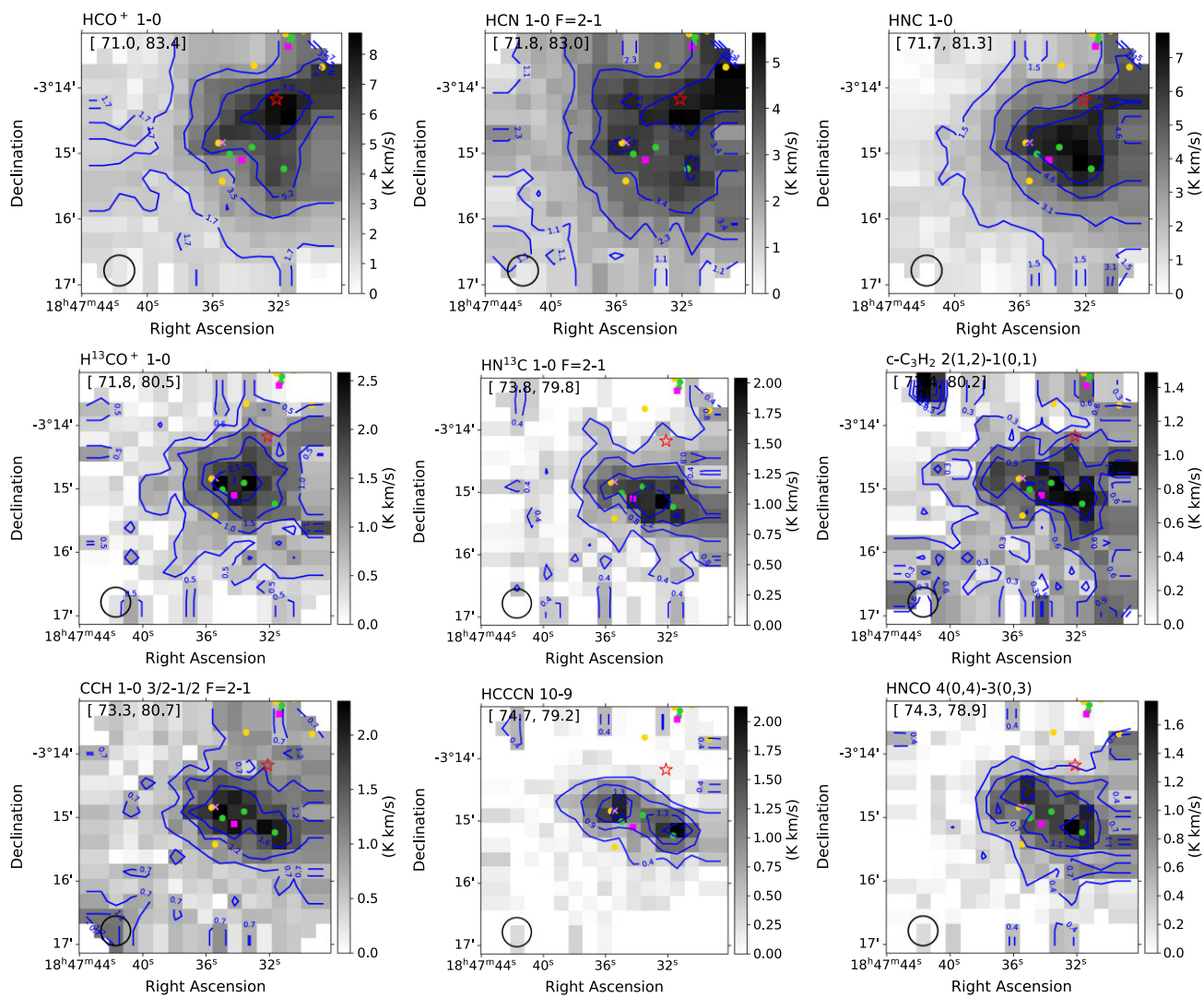

**Figure 2.** Integrated intensity maps of molecular lines observed in G029.60-0.63. The pentagram symbol denotes the position of identified infall source, the filled magenta squares denote the IRAS sources, and green and yellow points denote the Class 0/I and Class II YSOs from ALLWISE data, respectively. The crosses denote the maser sources.

The SiO (2-1) emission line is observed in G053.13+0.09, G081.72+0.57, G107.50+4.47, and G121.31+0.64. This line is often used to effectively trace outflow driven by YSOs. Among them, the SiO emission regions of G081.72+0.57 and G121.31+0.64 are relatively close to the positions of clump centre and the positions where the infall line profiles are significant. The SiO emission area of G053.13+0.09 is slightly east of the infall position, while the SiO of G107.50+4.47 is mainly distributed in clump A and also has a small amount in clump C. In addition, Yang et al. (2021) found that the $HCO^+$ lines of these four sources and the other two sources (G193.01+0.14 and G217.30-0.05) have red- and blue-line wing components. The integrated intensity maps of their line wings show bipolar outflow. The $HCO^+$ line of G029.60-0.63 has a blue-line wing component, which indicates that this source may have a unipolar outflow. G029.60-0.63, G053.13+0.09, G081.72+0.57, and G121.31+0.64 have been detected 6.7 GHz methanol masers and water masers (e.g. Genzel & Downes 1977; Szymczak, Hrynek & Kus 2000; Valdettaro et al. 2001; Pestalozzi, Minier & Booth 2005; De Villiers et al. 2014; Svoboda et al. 2016), while G107.50+4.47 and G217.30-0.05 have been detected water masers (e.g. Valdettaro et al. 2001), also suggesting that these sources have molecular outflows. No evidence of molecular outflow has been found in G082.21-1.53 and G109.00+2.73.

However, due to the limitations of the telescope's angular resolution, the more detailed spatial structures and properties of the clumps remain uncertain. In follow-up studies, mapping these infall sources with higher spatial resolution and sensitivity will help us answer these questions.

## 4 DISCUSSION

### 4.1 Physical parameters of the clumps

#### 4.1.1 Rotation temperature and column density of molecules

For some molecules with multiple transition lines or hyperfine structures, we can fit their spectral lines using the radiation transfer model to further limit their physical parameters such as excitation temperature and column density. Using XCLASS's local thermodynamic equilibrium (LTE) radiation transmission calculations can generate such a synthesized spectrum, and matched with our observation results. First, we need to exclude multipeaked line components that caused by optically thick lines' self-absorption. The central radial velocity of the clump is traced by an optically thin line. If the optically thick line has multipeaked components at the corresponding velocity, this line profile is likely to be caused by self-absorption. Since our sources are all confirmed infall sources, the optically thick lines, which are generally used to trace infall, such as HNC, HCO$^+$, and HCN, have complex spectral line profiles, and cannot be fitted by this method.

**Table 3.** Molecular rotation temperature and column density of sources.

| $T_{rot}$ | G029.60-0.63 (K) | G053.13+0.09 (K) | G081.72+0.57 (K) | G082.21-1.53 (K) | G107.50+4.47 (K) | G109.00+2.73 (K) | G121.31+0.64 (K) | G193.01+0.14 (K) | G217.30-0.05 (K) |
|---|---|---|---|---|---|---|---|---|---|
| $HC_5N$ | | | $75.6^{+4.72}_{-20.0}$ | | | | | | |
| $c-C_3H_2$ | $7.69^{+5.33}_{-2.19}$ | $9.21^{+6.09}_{-3.25}$ | $10.4^{+3.18}_{-1.75}$ | $11.4^{+8.49}_{-4.55}$ | $9.05^{+9.66}_{-3.75}$ | $6.19^{+10.4}_{-1.19}$ | $9.74^{+9.49}_{-3.66}$ | $8.16^{+9.78}_{-3.16}$ | $7.30^{+9.92}_{-2.99}$ |
| $CH_3CCH$ | | $16.7^{+6.23}_{-4.16}$ | $29.9^{+5.94}_{-7.61}$ | | | | $16.8^{+5.31}_{-4.01}$ | | |
| $C_4H$ | | | $78.1^{+31.0}_{-22.9}$ | $105^{+24.9}_{-38.3}$ | | | | | |
| SO | | $38.2^{+21.5}_{-10.9}$ | $30.9^{+19.5}_{-9.06}$ | | $29.6^{+20.8}_{-12.8}$ | $23.2^{+6.16}_{-9.17}$ | $23.4^{+20.6}_{-9.93}$ | $35.8^{+4.11}_{-7.96}$ | $12.3^{+7.75}_{-4.35}$ |
| CCS | | | $16.3^{+20.0}_{-7.63}$ | | $10.8^{+7.94}_{-3.33}$ | | | | |
| HCO | | | $113^{+9.10}_{-22.1}$ | | | $61.2^{+8.10}_{-25.2}$ | | | |
| CCH | $35.5^{+14.9}_{-12.9}$ | $47.5^{+24.3}_{-17.4}$ | $40.4^{+7.66}_{-6.43}$ | $31.1^{+17.6}_{-7.30}$ | $44.1^{+13.3}_{-15.3}$ | $33.8^{+14.6}_{-11.5}$ | $52.1^{+7.39}_{-37.6}$ | $34.1^{+14.6}_{-11.6}$ | $36.2^{+14.5}_{-13.4}$ |
| HNCO | $12.0^{+2.45}_{-6.94}$ | $6.92^{+2.34}_{-1.60}$ | $9.16^{+9.72}_{-2.91}$ | | $10.69^{+10.0}_{-2.57}$ | $10.0^{+5.00}_{-3.14}$ | $10.3^{+9.85}_{-2.65}$ | | |
| HCCCN | $49.6^{+50.6}_{-28.9}$ | $171^{+18.9}_{-58.5}$ | $280^{+39.6}_{-24.9}$ | $71.4^{+57.2}_{-35.8}$ | $47.7^{+118}_{-33.0}$ | | $73.3^{+38.5}_{-38.6}$ | $62.2^{+43.5}_{-30.1}$ | |
| $CH_2CO$ | | | $59.1^{+11.1}_{-18.1}$ | | | | | | |
| $H_2CS$ | | $12.9^{+22.3}_{-3.90}$ | $23.6^{+30.1}_{-2.35}$ | | $27.0^{+8.82}_{-7.16}$ | $33.9^{+38.7}_{-13.5}$ | $16.7^{+21.5}_{-5.30}$ | $53.6^{+12.0}_{-19.5}$ | |
| $SO_2$ | | | $26.5^{+15.1}_{-10.7}$ | | $13.7^{+19.4}_{-5.25}$ | $12.3^{+5.31}_{-2.14}$ | | | |
| $N_{mole}$ | ($\times 10^{12}$ cm$^{-2}$) | ($\times 10^{12}$ cm$^{-2}$) | ($\times 10^{12}$ cm$^{-2}$) | ($\times 10^{12}$ cm$^{-2}$) | ($\times 10^{12}$ cm$^{-2}$) | ($\times 10^{12}$ cm$^{-2}$) | ($\times 10^{12}$ cm$^{-2}$) | ($\times 10^{12}$ cm$^{-2}$) | ($\times 10^{12}$ cm$^{-2}$) |
| $HC_5N$ | | | $1.95^{+0.27}_{-0.39}$ | | | | | | |
| $c-C_3H_2$ | $3.01^{+1.00}_{-0.22}$ | $1.49^{+0.31}_{-0.37}$ | $11.6^{+1.53}_{-4.97}$ | $1.55^{+1.51}_{-0.18}$ | $1.95^{+0.27}_{-0.39}$ | $0.92^{+0.37}_{-0.32}$ | $3.66^{+0.13}_{-0.62}$ | $2.31^{+1.33}_{-0.28}$ | $1.46^{+0.33}_{-0.35}$ |
| $CH_3CCH$ | | $4.93^{+0.37}_{-0.34}$ | $212^{+9.35}_{-25.3}$ | | | | $21.5^{+1.07}_{-1.16}$ | | |
| $C_4H$ | | | $55.4^{+10.3}_{-2.02}$ | $32.0^{+1.60}_{-5.29}$ | | | | | |
| SO | | $9.37^{+0.38}_{-0.33}$ | $125^{+14.9}_{-50.0}$ | | $22.7^{+2.46}_{-4.18}$ | $4.92^{+0.49}_{-0.21}$ | $9.81^{+2.29}_{-1.03}$ | $15.8^{+3.15}_{-3.58}$ | $7.33^{+0.37}_{-0.34}$ |
| CCS | | | $2.12^{+0.27}_{-0.39}$ | | $0.48^{+0.09}_{-0.07}$ | | | | |
| HCO | | | $239^{+23.8}_{-42.6}$ | | | $25.5^{+2.31}_{-4.35}$ | | | |
| CCH | $104^{+15.1}_{-50.0}$ | $82.8^{+10.8}_{-1.94}$ | $1420^{+68.4}_{-197}$ | $49.8^{+1.52}_{-4.23}$ | $119^{+11.2}_{-48.5}$ | $62.2^{+1.45}_{-4.48}$ | $180^{+42.2}_{-31.3}$ | $90.0^{+1.43}_{-4.99}$ | $94.4^{+1.44}_{-5.01}$ |
| HNCO | $2.59^{+0.19}_{-0.57}$ | $0.49^{+0.04}_{-0.03}$ | $5.18^{+0.10}_{-0.22}$ | | $0.64^{+0.04}_{-0.03}$ | $0.31^{+0.05}_{-0.03}$ | $1.19^{+0.35}_{-0.33}$ | | |
| HCCCN | $0.45^{+0.78}_{-0.20}$ | $0.74^{+0.04}_{-0.03}$ | $128^{+8.48}_{-16.4}$ | $0.32^{+0.17}_{-0.05}$ | $0.40^{+0.03}_{-0.02}$ | | $1.58^{+0.53}_{-0.28}$ | $0.28^{+0.04}_{-0.02}$ | |
| $CH_2CO$ | | | $22.2^{+2.60}_{-4.04}$ | | | | | | |
| $H_2CS$ | | $1.24^{+0.35}_{-0.35}$ | $51.6^{+0.98}_{-2.90}$ | | $7.09^{+0.99}_{-0.22}$ | $0.97^{+0.04}_{-0.03}$ | $5.65^{+0.91}_{-0.23}$ | $5.47^{+0.35}_{-0.33}$ | |
| $SO_2$ | | | $41.6^{+1.36}_{-4.85}$ | | $7.71^{+0.31}_{-0.37}$ | $1.15^{+0.42}_{-0.40}$ | | | |

We only fit those molecular lines with high-signal-to-noise ratio and multiple transitions in the frequency range we observed, such as c-C$_3$H$_2$, CCH, HNCO, and HC$_3$N. The parameters required for XCLASS are the clump size, the rotation temperature, the column density, the full width at half-maximum (FWHM) of the spectral lines, and the velocity offset relative to the systemic velocity of the clump. Among them, the clump size is obtained by a 2D Gaussian fitting to the integrated intensity map of this line. The velocity offset and line width are estimated by Gaussian fitting of the spectral lines. We assume that the rotation temperature of the molecule is 20 K, and the molecular column density is about $1 \times 10^{13}$ cm$^{-2}$. Different optimization algorithms in Modelling and Analysis Generic Interface for eXternal numerical codes (MAGIX, Möller et al. 2013) can be used to explore the parameter space and minimize the distribution space. The fitting algorithms used here are 'genetic', 'Levenberg–Marquardt' and 'errorestim-ins'. We keep modifying the parameters variation range until the spectral lines generated by the model match the observed lines. The HN$_2$D and HCN isotope molecules were not fitted because the frequency resolution of the observations could not well distinguish the hyperfine structures. For other molecules with only one observed transition line, the temperature and column density obtained from the multitransition line molecules are used as initial values, and also fitted the spectral lines with MAGIX to obtain the fitting results. However, the uncertainty of the obtained results is relatively large, especially the molecular rotation temperature. We have not adopted the results for these molecules with only one transition line. Therefore, only the rotation temperatures and column densities of 13 molecules are listed in Table 3 (the FWHM and the corresponding centre radial velocity of all molecules are listed in Tables C1 and C2, respectively).

We also tried to estimate the spatial distribution of these parameters over the entire observation area. Based on the results obtained from the average spectrum, we use them as initial parameters to fit the data for each pixel in the map. Unfortunately, XCLASS does not constrain the molecular rotation temperatures well, and the final results are greatly affected by the initial values and the range of parameter variation. The error values may even be several times larger than the results, which makes it difficult to determine the reliability of the estimated results given by the software. The reason for this may be due to the insufficient signal-to-noise ratio or the self-absorption of some molecular lines, which makes the software unable to fit the molecular emission well at each pixel.

**Table 4.** Observed and simulated average fractional abundances of 13 species with respect to H$_2$ in nine infall candidates.

| Species | G029.60-0.63 | | G053.13+0.09 | | G081.72+0.57 | | G082.21-1.53 | | G107.50+4.47 | |
|---|---|---|---|---|---|---|---|---|---|---|
| | Observation | Simulation | Observation | Simulation | Observation | Simulation | Observation | Simulation | Observation | Simulation |
| $N(H_2)$ (cm$^{-2}$) | 3.98(22)[b] | | 1.00(22) | | 1.58(23) | | 1.00(22) | | 2.51(22) | |
| CCH | 2.61(−09) | 2.49(−08) | 8.28(−09) | 2.53(−08) | 8.99(−09) | 2.67(−08) | 4.98(−09) | 2.55(−08) | 4.74(−09) | 2.65(−08) |
| C$_4$H | | | | | 3.51(−10) | 2.69(−09) | 3.20(−09) | 2.33(−09) | | |
| C$_3$H$_2$[a] | 1.01(−10) | 1.03(−10) | 1.99(−10) | 1.27(−10) | 9.79(−11) | 1.57(−10) | 2.07(−10) | 1.02(−10) | 1.04(−10) | 1.54(−10) |
| CH$_3$CCH | | | 4.93(−10) | 5.44(−11) | 1.34(−09) | **7.00(−11)** | | | | |
| HC$_3$N | 1.13(−11) | 4.48(−11) | 7.40(−11) | 7.49(−11) | 8.10(−10) | 1.22(−10) | 3.20(−11) | 4.43(−11) | 1.59(−11) | 1.17(−10) |
| HC$_5$N | | | | | 1.23(−11) | 1.22(−11) | | | | |
| CCS | | | | | 1.34(−10) | **2.75(−09)** | | | 1.91(−11) | **2.73(−09)** |
| CH$_2$CO | | | | | 1.41(−10) | 1.01(−09) | | | | |
| H$_2$CS | | | 1.24(−10) | **1.95(−09)** | 3.27(−10) | 2.42(−09) | | | 2.82(−10) | 2.37(−09) |
| HCO | | | | | 1.51(−09) | **1.16(−10)** | | | | |
| HNCO | 6.51(−11) | **2.26(−13)**[c] | 4.90(−11) | **4.77(−13)** | 3.28(−11) | **8.33(−13)** | | | 2.55(−11) | **7.91(−13)** |
| SO | | | 9.37(−10) | 8.61(−10) | 7.91(−10) | 9.84(−10) | | | 9.04(−10) | 9.67(−10) |
| SO$_2$ | | | | | 2.63(−10) | 3.21(−11) | | | 3.07(−10) | 3.11(−11) |
| Number of fits[d] | 4 | 3 | 7 | 5 | 13 | 9 | 4 | 4 | 8 | 6 |
| $\kappa$[e] | | 0.471 | | 0.579 | | 0.455 | | 0.753 | | 0.440 |
| $t$[f] (yr) | | 1.68(04) | | 1.77(04) | | 1.91(04) | | 1.67(04) | | 1.89(04) |

| Species | G109.00+2.73 | | G121.31+0.64 | | G193.01+0.14 | | G217.30-0.05 | |
|---|---|---|---|---|---|---|---|---|
| | Observation | Simulation | Observation | Simulation | Observation | Simulation | Observation | Simulation |
| $N(H_2)$ (cm$^{-2}$) | 2.51(22) | | 1.58(22) | | 1.58(22) | | 2.51(22) | |
| CCH | 2.48(−09) | **2.71(−08)** | 1.14(−08) | 2.98(−08) | 5.70(−09) | 2.55(−08) | 3.76(−09) | 2.49(−08) |
| C$_4$H | | | | | | | | |
| C$_3$H$_2$ | 4.89(−11) | 1.65(−10) | 3.09(−10) | 2.28(−10) | 1.95(−10) | 1.02(−10) | 7.76(−11) | 1.03(−10) |
| CH$_3$CCH | | | 1.36(−09) | 1.37(−10) | | | | |
| HC$_3$N | | | 1.00(−10) | 1.76(−10) | 1.77(−11) | 4.43(−11) | | |
| HC$_5$N | | | | | | | | |
| CCS | | | | | | | | |
| CH$_2$CO | | | | | | | | |
| H$_2$CS | 3.60(−11) | **2.53(−09)** | 3.58(−10) | 3.30(−09) | 3.46(−10) | 1.36(−09) | | |
| HCO | 1.02(−09) | 1.18(−10) | | | | | | |
| HNCO | 1.24(−11) | **9.25(−13)** | 7.53(−11) | **8.25(−13)** | | | | |
| SO | 1.96(−10) | 1.03(−09) | 6.21(−10) | 1.80(−09) | 9.97(−10) | 7.64(−10) | 2.92(−10) | 7.63(−10) |
| SO$_2$ | 4.58(−11) | 3.37(−11) | | | | | | |
| Number of fits | 7 | 4 | 7 | 6 | 5 | 5 | 3 | 3 |
| $\kappa$ | | 0.419 | | 0.532 | | 0.689 | | 0.663 |
| $t$ (yr) | | 1.96(04) | | 2.54(04) | | 1.67(04) | | 1.68(04) |

[a]The observed abundances of C$_3$H$_2$ represent the observations of c-C$_3$H$_2$, while the simulated abundances of C$_3$H$_2$ represent the sum abundances of two isomers, c-C$_3$H$_2$ and l-C$_3$H$_2$.
[b]$a(b) = a \times 10^b$.
[c]Bold type indicates overproduction compared with the upper limit or underproduction compared with the lower limit, which cannot fit observations.
[d]Number of fits represents how many species can fit observations at $t$.
[e]$\kappa$ is the maximum average confidence level of all observed species in each source.
[f]$t$ is the time-scale from the IRDC stage when $\kappa$ reaches the maximum value.

*4.1.2 Abundances*

In Yang et al. (2020), the C$^{18}$O data from the MWISP project was used to estimate the column density of H$_2$. The abundance ratio of C$^{18}$O to H$_2$ is assumed to be $N(H_2)/N(C^{18}O) = 7 \times 10^6$ (e.g. Castets & Langer 1995; Warin et al. 1996). The H$_2$ column densities of these nine sources are all between $10^{22}$ cm$^{-2}$ and $10^{23}$ cm$^{-2}$, which are relatively dense molecular clumps. However, there are some uncertainties in the H$_2$ column density estimated in this way, such as the assumption of local thermal equilibrium, the uncertainties of $T_{ex}$ and N(H$_2$)/N(C$^{18}$O) abundance ratio. All these factors may cause some errors in our estimated H$_2$ column densities, but in most cases the errors will not be over an order of magnitude. We use the estimated H$_2$ column densities to calculate the abundance of each observed molecule ($X(mole) = N(mole)/N(H_2)$). These abundances are listed in Table 4. Among them, the abundance of C$_3$H$_2$ only represents that the observed value of c-C$_3$H$_2$ with the ortho-para ratio is about 3 (e.g. Turner, Lee & Herbst 1998; Takakuwa et al. 2001).

**4.2 Modelling observed chemical abundances**

To investigate the chemical evolution of these nine infall candidates, we simulate chemical models and analyse the agreement between them and observations. The simulations are calculated via the astrochemical code Nautilus (Ruaud, Wakelam & Hersant 2016), and for simplicity, the two-phase gas–grain model is adopted (Hasegawa et al. 1992). The reaction network is the same as the one provided by Wang et al. (2021), which includes the photodesorption mechanism with the external UV radiation scaling factor $G_0 = 1$, the UV radiation scaling factor induced by the cosmic ray particles $G'_0 = 10^{-4}$, and the interstellar UV radiation field $F_0 = 10^8$ photons cm$^{-2}$ s$^{-1}$ (Öberg, van Dishoeck & Linnartz 2009a; Öberg et al. 2009b; Chang & Herbst 2014). The reactive desorption efficiency $a_{RRK}$ is 0.01 (Garrod, Wakelam & Herbst 2007), the sticking coefficient $P_s$ is set to be 1, and the ratio of the diffusion barrier $E_b$ to the desorption energy $E_d$ is set to be 0.5. Other parameters are the same as in Table 1 listed by Wang, Chang & Wang (2019) except for a higher cosmic ray ionization rate $\zeta_{CR} = 1.8 \times 10^{-16}$ s$^{-1}$ based on a study of multiple high-mass star-forming regions (HMSFRs; Indriolo et al. 2015; Gieser et al. 2021). Considering that these nine infall candidates belong to intermediate or massive star-forming regions, not only the initial abundances provided by Gerner et al. (2014) are adopted, but also the physical model. It is assumed to be a spherically symmetric structure and includes four evolutionary stages, which are the infrared dark cloud (IRDC) stage, the high-mass protostellar object (HMPO) stage, the hot molecular core (HMC) stage, and the ultra-compact HII (UCHII) stage, respectively. The radial hydrogen density $n_{H_2}$ and the temperature $T$ distribution both follow power laws from the centre. Therefore, the chemical evolutions at different locations from the centre ($r_{out} \sim 0.5$ pc as the outermost region for dense parts of HMSFRs) are calculated, then the average abundances among an HMSFR can be acquired for comparing with observations. However, the time-scales of four evolutionary stages suggested by Gerner et al. (2015) are utilized, which are $1.65 \times 10^4$ yr, $3.2 \times 10^4$ yr, $3.5 \times 10^4$ yr, and $3 \times 10^3$ yr, respectively, since the time-scales provided by Gerner et al. (2014) cannot fit the observations well, neither the model used by Gerner et al. (2015). We calculate the evolutions in $10^5$ yr for simplicity, thus the time-scale of the UCHII stage is set to be $1.65 \times 10^4$ yr, which is still consistent with the range in Gerner et al. (2014 and 2015). In addition, since these nine sources still evolve at very early stages, the simulations in IRDC and HMPO stage are mainly compared and discussed.

Fig. 3 depicts the evolution of the average fractional abundances of 13 observed species with respect to $H_2$ among $10^5$ au ($\sim 0.5$ pc) around the HMSFR centre. To clearly show the chemical evolution in different stages, the left-hand panels present the evolution in the IRDC stage within $1.65 \times 10^4$ yr, while the right-hand panels present the evolution in the HMPO, HMC, and UCHII stage within $1.65 \times 10^4$–$10^5$ yr. Fig. 4 depicts the different evolutions of these 13 species at $10^2$, $10^3$, $10^4$, and $10^5$ au from the centre. Overall, the abundance of each species can range in several orders of magnitude from the location at $10^2$–$10^5$ au from the centre, since $n_{H_2}$ and $T$ deviate significantly from inner regions to outer regions. In the IRDC stage, most species increase to reach peak average abundances at $t \sim 10^2$–$10^3$ yr then decrease, except that the abundances of CH$_3$CCH, CCS, H$_2$CS, HNCO, and SO$_2$ increase monotonically. During the early cold stage, gas-phase reactions dominate for producing carbon-chain species and simple molecules thus induce the increase of abundances, then the adsorption overwhelms to induce most molecules to be accreted on the grains and decrease the abundances. Higher densities can accelerate the adsorption in shorter time. For example, in Fig. 4, all 13 abundances at $10^2$ au from the centre decrease dramatically when $t < 100$ yr with $n_{H_2} = 2 \times 10^9$ cm$^{-3}$, while they decrease later when $t \sim 300 - 1000$ yr with $n_{H_2} = 1.2 \times 10^7$ cm$^{-3}$ at $10^3$ au from the centre. Thus in the outer regions with lower densities, the time-scale of $1.65 \times 10^4$ yr is not long enough to decrease to abundances via adsorption, such as $n_{H_2} = 5 \times 10^3$ cm$^{-3}$ at $10^5$ au from the centre. For CCH, C$_4$H, C$_3$H$_2$, HC$_3$N, HC$_5$N, CH$_2$CO, HCO, and SO, rapid synthesis and adsorption generate peak abundances. On the other hand, for CH$_3$CCH, CCS, H$_2$CS, HNCO, and SO$_2$, less efficient synthesis cannot produce enough molecules to create peak values. In the next HMPO stage, most average abundances increase within one order of magnitude, mainly because of the thermal desorption of molecules on the grains with higher temperatures than the IRDC stage, especially at the start of HMPO stage when $t < 2 \times 10^4$ yr for the regions within $10^3$ au around the centre with $T \sim 77$ K. However, for the evolution at $10^4$ au around the centre with $T \sim 32$ K, the increase of seven carbon-chain species corresponds to the warm-carbon chain chemistry (WCCC) induced by the sublimation of CH$_4$ on the grains (Hassel, Herbst & Garrod 2008; Sakai et al. 2008, 2009; Hassel, Harada & Herbst 2011; Wang et al. 2019) and subsequent gas-phase reactions. For the outermost region at $10^5$ au around the centre, the evolution is similar to that in the IRDC stage since $n_{H_2}$ and $T$ change a little. Finally, the average abundances in the HMC and UCHII stage still fluctuate within one order of magnitude. For different locations, the rapid increase still corresponds to the thermal desorption with high temperatures, and the decrease is induced by more efficient gas-phase reactions.

To compare the simulations with observations quantitatively and analyse probable ages of these sources, the 'mean confidence level' $\kappa$ (Garrod et al. 2007; Hassel et al. 2008; Mookerjea et al. 2012; Wang et al. 2019, 2021) for each source is calculated. Table 4 presents the observed and simulated average abundances of 13 species with respect to $H_2$ in these nine sources. Two best fits can be acquired in the IRDC stage and HMPO stage, respectively, but the time-scale of the former one is only between several hundred to several thousand years, which seems to be too short and difficultly observed but may be a lower limit as suggested by Gerner et al. (2015). Therefore, only the best fit in the HMPO stage for each source is listed in Table 4, which makes most species fit the observations. In eight sources, the best fit occurs at the start of the HMPO stage with $t \sim 1.65 - 2 \times 10^4$ yr, only $t \sim 2.54 \times 10^4$ yr in G121.31+0.64. The chemical evolution of these sources is similar to WCCC sources L1527 and B228 (Sakai et al. 2008, 2009), especially in the inner $2 \times 10^4$ au region with $T > 25$ K and $n_{H_2} > 6.7 \times 10^4$ cm$^{-3}$. Such temperature of G081.72+0.57 or DR21(OH) is also consistent with the analysis presented by Mookerjea et al. (2012). The best-fitting numbers of species for most sources can last several thousand years at $t < 3 \times 10^4$ yr with smaller values of $\kappa$. However, the time-scales of different evolutionary stages in HMSFRs are not clearly confirmed with an uncertainty of a factor of 2–3 (Gerner et al. 2014), also $n_{H_2}$ and $T$ can be very different, so the best fit possibly occurs at different time-scales adopting other physical models. Nevertheless, it is reliable that these sources evolve in the early HMPO stage when the temperature in the inner region reaches several 10 K ($t \sim 1.65 - 2 \times 10^4$ yr in this work).

According to the comparison, most species can fit the observations, but CH$_3$CCH and HCO, and especially HNCO are underproduced by over one order of magnitude in the simulations. On the contrary, CCS and H$_2$CS are overproduced by over one order of magnitude. The discrepancies between simulations and observations are also intuitively depicted in Fig. 5. Although some average abundances cannot fit observations as mentioned before, the abundances in different locations around the centre at the same time deviate from observations within one order of magnitude as shown in Fig. 5, including CH$_3$CCH, H$_2$CS, HCO, and HNCO. It may indicate that these species could be produced in certain physical environment in such complex HMSFRs. Additional higher sensitivity and resolution observations may confirm whether these species distribute at different locations as simulations. Another possible cause for a discrepancy between model and data could be that some of the real sources have multiple clumps, while the model assumed a single-source centre. On the other hand, the reaction network may be not complete for producing CH$_3$CCH, HCO, and HNCO, meanwhile destructing CCS and H$_2$CS. In addition, the coefficients of related reactions may be not accurate to influence their abundances. Thus, a more accurate reaction network needs to be proposed by astrochemists. Furthermore, a more detailed physical model for HMSFRs is necessary in the future simulations.

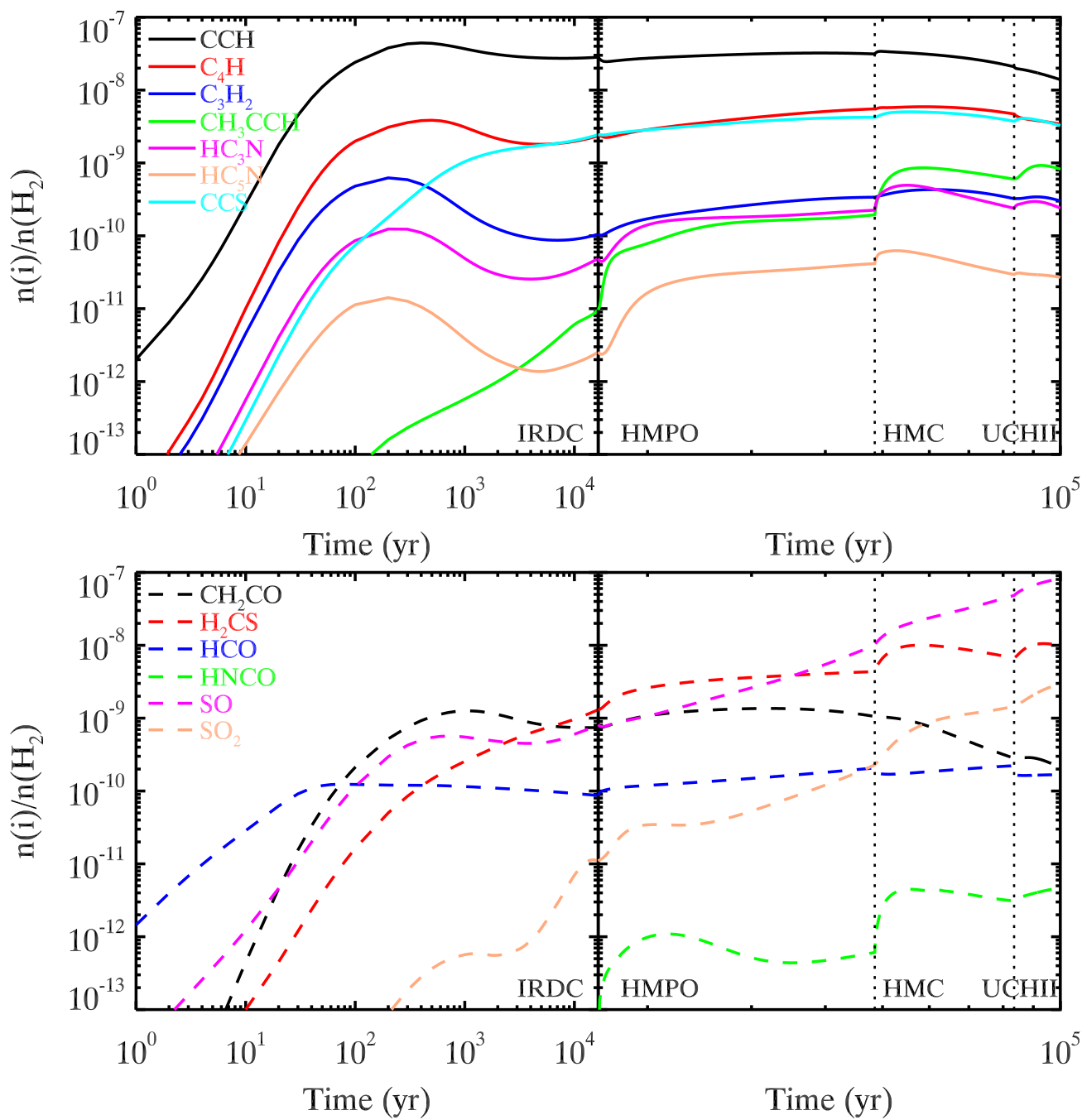

**Figure 3.** Time-dependent evolution of the average fractional abundances of 13 observed species among $10^5$ au around the HMSFR centre. The left-hand panels depict the chemical evolution in the IRDC stage, while the right-hand panels depict the chemical evolution in the HMPO, HMC, and UCHII stage.

## 5 CONCLUSION

We analysed nine infall candidates with significant blue profiles observed by the FTS wide-sideband mode of the IRAM 30-m telescope. These sources are identified in Yang et al. (2021). The frequency ranges we observed are 83.7–91.5 GHz and 99.4–107.2 GHz. We observed multiple molecular lines in these sources, and used these multiline data to analyse the physical and chemical properties of the infall candidates. The conclusions are as follows:

(i) We used XCLASS to identify the observed spectral lines, and identified about 7–27 kinds of different molecules and isotopes emission lines in nine infall candidates. The emission lines of H$^{13}$CO$^+$, HCO$^+$, HCN, HNC, c-C$_3$H$_2$, and CCH molecules were observed in all sources. The source with the fewest observed molecular species is G217.30-0.05. The sources associated with the massive star-forming

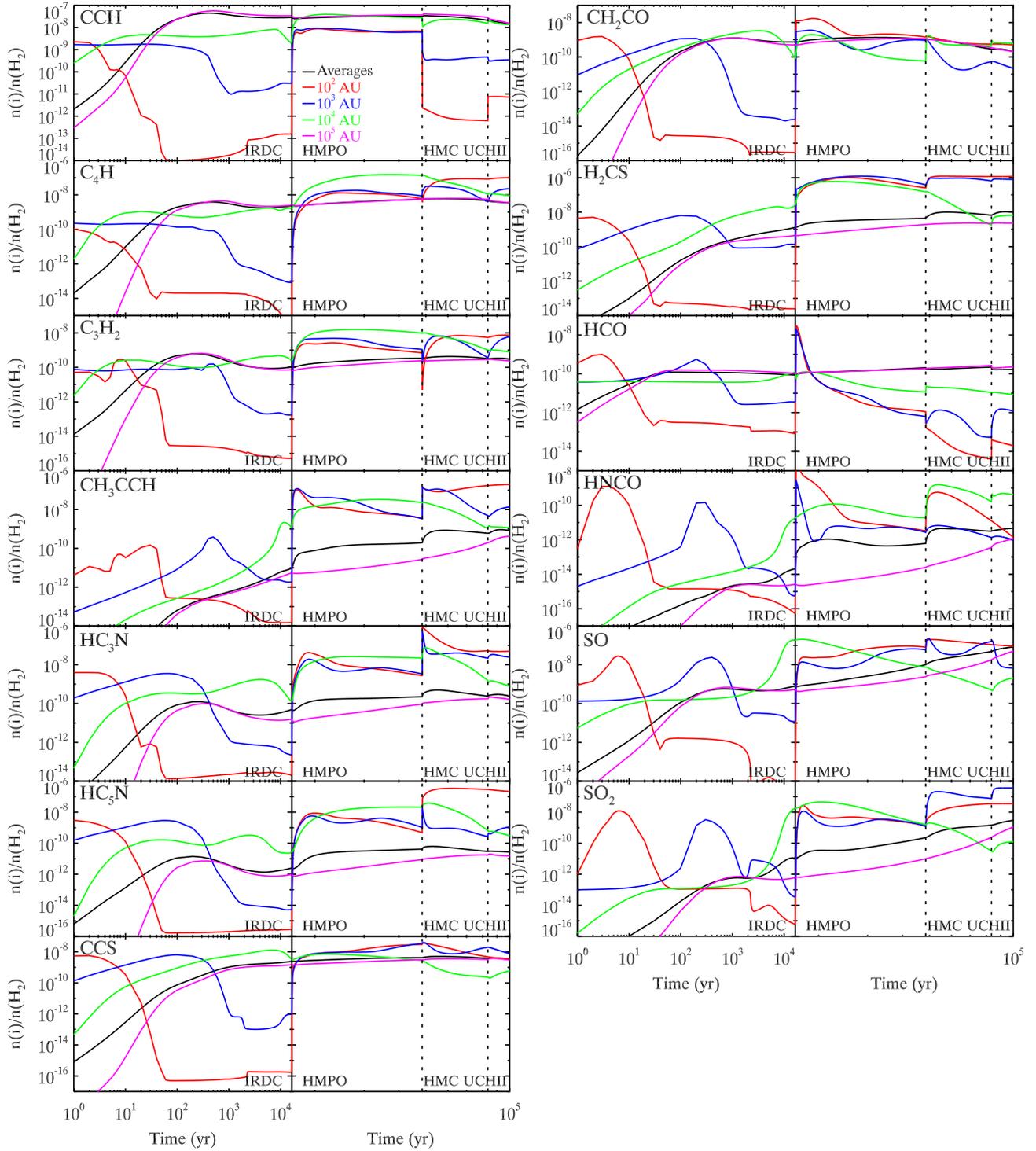

**Figure 4.** Time-dependent evolution of the fractional abundances of 13 observed species at different location ($10^2$ au, $10^3$ au, $10^4$ au, and $10^5$ au) from the HMSFR centre. The average fractional abundances among $10^5$ au around the centre, which are the same as in Fig. 3, are also plotted.

regions have more kinds of molecular species, such as G081.72+0.57 and G121.31+0.64.

(ii) In most cases, the integrated intensity maps of different molecular lines from the same source will display spatially correlated compact regions, and are usually associated with IR point sources, YSOs, and/or masers. However, in some sources, there are some molecular emissions that have slightly different spatial distributions. In addition, SiO (2-1) was observed in four sources, i.e.

G053.13+0.09, G081.72+0.57, G107.50+4.47, and G121.31+0.64, suggesting that these sources may have molecular outflow, and their HCO$^+$ (1-0) line wings also confirm it.

(iii) We used XCLASS's local LTE radiation transmission calculations to estimate the molecular rotation temperatures and column densities, and compare the molecular column densities with the H$_2$ column density, resulting the abundance of each molecule in these nine infall candidates.

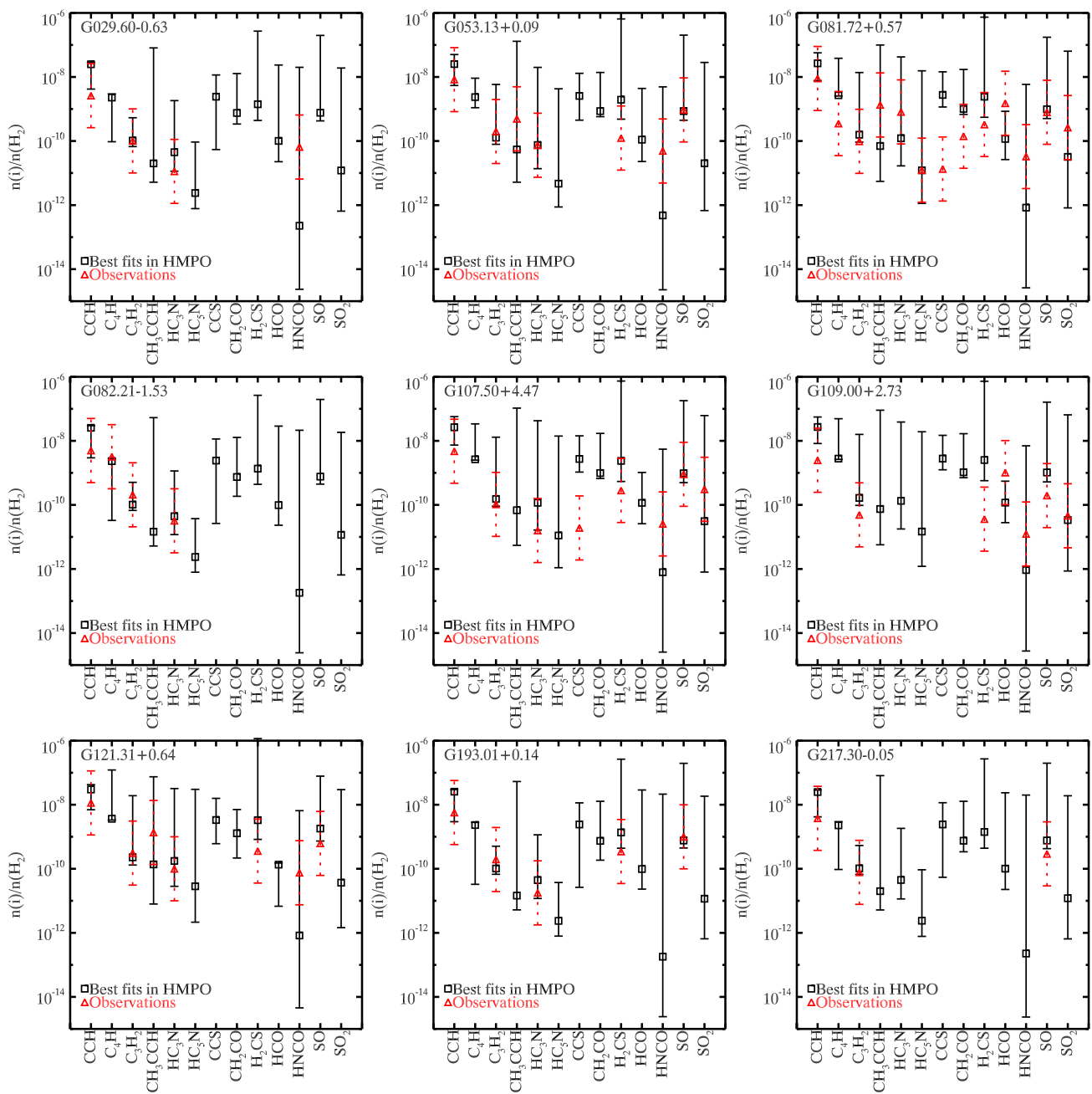

**Figure 5.** Comparison of simulated abundances at the best-fitting time in HMPO stage and observations for each source. The average abundances among $10^5$ au around the centre are labelled as squares while the observed abundances are labelled as triangles. The upper and lower limits of simulations represent the maximums and minimums at different locations in $10^5$ au around the centre at the certain time, respectively. The upper and lower limits of observations represent one order of magnitude higher and lower than observed abundances, respectively.

(iv) We compared the observed molecular abundances with those simulated by chemical models, and estimated the evolutionary stages and time-scales of these infall candidates. Since our sources belong to intermediate or massive star-forming regions, a physical model of HMSFRs was adopted. The results show that, for most sources, the best fit occurs at the beginning of the HMPO stage. But there are still some species that do not fit the observations very well. Higher angular resolution observations are needed to confirm the details of their spatial distribution.

**ACKNOWLEDGEMENTS**

We are grateful to the staff of the Institut de Radio astronomie Millimétrique (IRAM) for their assistance and support during the observations. Thanks to Prof. Sheng-Li Qin of Yunnan University for his help in the identification and analysis of spectral lines. This work is supported by the National Key R&D Program of China (grant no. 2017YFA0402702), and the National Natural Science Foundation of China (NSFC; grant nos. 11873093 and U2031202). YW acknowledges the support by the NSFC grant no. 11973090, 11873094, 12041305, and the Natural Science Foundation of Jiangsu Province (grants no. BK20221163). ZC acknowledges the support from the NSFC general grant 11903083.

## DATA AVAILABILITY

The data underlying this article are available in the article and in the IRAM Science Data Archive (project code:134-18 & 032-19).

## SUPPORTING INFORMATION

Supplementary data are available at *MNRAS* online.

**mnras_R2_v2_add.pdf**

Please note: Oxford University Press is not responsible for the content or functionality of any supporting materials supplied by the authors. Any queries (other than missing material) should be directed to the corresponding author for the article.

This paper has been typeset from a TeX/LaTeX file prepared by the author.